%% file: main.tex
\documentclass[10pt,twocolumn,letterpaper]{article}

\usepackage[pagenumbers]{cvpr} 
\input{preamble}

\definecolor{cvprblue}{rgb}{0.21,0.49,0.74}
\usepackage[pagebackref,breaklinks,colorlinks,citecolor=cvprblue]{hyperref}
\usepackage{makecell}
\usepackage{float}

\def\confName{CVPR}
\def\confYear{2025}

\title{GroomLight: Hybrid Inverse Rendering for\\Relightable Human Hair Appearance Modeling}

\author{Yang Zheng\textsuperscript{1,2} \quad Menglei Chai\textsuperscript{2} \quad Delio Vicini\textsuperscript{2} \quad Yuxiao Zhou\textsuperscript{2,3} \quad Yinghao Xu\textsuperscript{1} \\ \\ 
\quad Leonidas Guibas\textsuperscript{1} \quad Gordon Wetzstein\textsuperscript{1} \quad Thabo Beeler\textsuperscript{2} \\ \\
\textsuperscript{1}Stanford University \quad  \textsuperscript{2}Google \quad \textsuperscript{3}ETH Zurich
}
\begin{document}

\twocolumn[{%
\renewcommand\twocolumn[1][]{#1}%
\maketitle

\begin{center}
    \centering
    \captionsetup{type=figure}
    \includegraphics[width=0.99\linewidth]{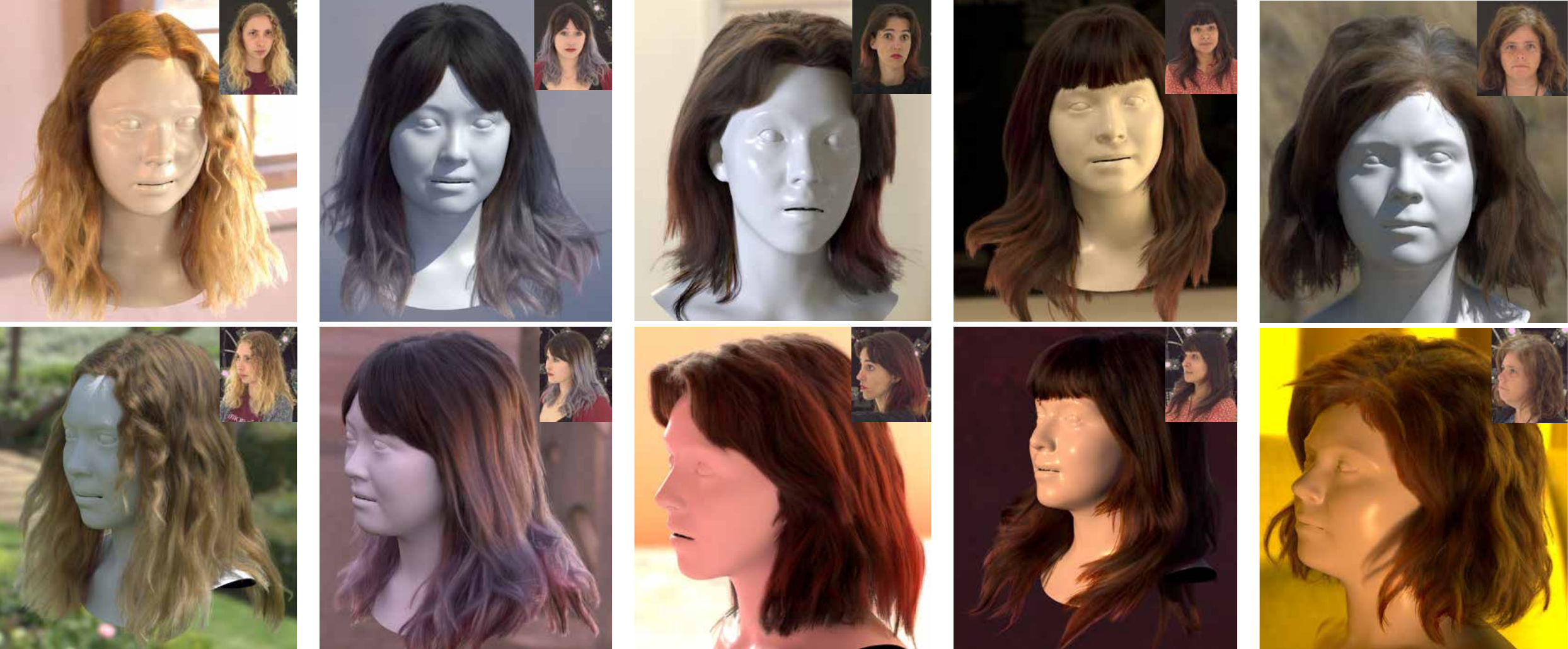}
    \captionof{figure}{\textit{GroomLight} achieves high-fidelity reconstruction of human hair appearance from real-world images, enabling realistic rendering under diverse lighting conditions. Here, in each column, we present relighting results of a same subject under two different environments, using the appearance model reconstructed by \textit{GroomLight}. One input view with the similar head pose is shown at the top right corner.} 
    \label{fig:teaser}
\end{center}
}]

\input{sec/0_abstract}    
\input{sec/1_intro}
\input{sec/2_related}
\input{sec/3_method}
\input{sec/4_experiments}
\input{sec/5_conclusion}
\input{sec/6_ack}

{
    \small
    \bibliographystyle{ieeenat_fullname}
    \bibliography{main}
} 

\appendix
\twocolumn[{%
\renewcommand\twocolumn[1][]{#1}%
\maketitle
\begin{center}
\textbf{\Large Supplementary Materials}
\end{center}
}]

\input{sec/suppl}

\end{document}

%% file: preamble.tex
\usepackage[dvipsnames]{xcolor}


%% file: sec/0_abstract.tex
\begin{abstract}

We present GroomLight, a novel method for relightable hair appearance modeling from multi-view images. Existing hair capture methods struggle to balance photorealistic rendering with relighting capabilities. Analytical material models, while physically grounded, often fail to fully capture appearance details. Conversely, neural rendering approaches excel at view synthesis but generalize poorly to novel lighting conditions. GroomLight addresses this challenge by combining the strengths of both paradigms. It employs an extended hair BSDF model to capture primary light transport and a light-aware residual model to reconstruct the remaining details. We further propose a hybrid inverse rendering pipeline to optimize both components, enabling high-fidelity relighting, view synthesis, and material editing. Extensive evaluations on real-world hair data demonstrate state-of-the-art performance of our method. Our project website is at: \url{https://syntec-research.github.io/GroomLight}.

\end{abstract}

%% file: sec/1_intro.tex
\section{Introduction}
\label{sec:intro}
Hair capture is a long-standing problem in computer graphics and vision, and plays an essential role in human digitization. While substantial progress has been made in strand-accurate geometry reconstruction, hair appearance modeling and photorealistic relighting remain open challenges.

As part of efforts towards physics-based photorealistic rendering, analytical material models for hair have been extensively studied. Following this path, inverse rendering approaches~\cite{sun2021human} seek to match renderings to input images by optimizing hair material properties, specifically the bidirectional scattering distribution function (BSDF). A key advantage of such approaches lies in their exceptional ability to generalize to novel viewpoints, unseen lighting, and altered hair geometry. 
However, even with sophisticated differentiable renderers, the reconstruction methods suffer from system errors stemming from factors such as imperfect geometry reconstruction, inaccurate lighting calibration, simplified light scattering models without diffraction and interference effects, etc.
Consequently, these methods struggle to reproduce critical appearance details, hindering their successful application to real images.

Meanwhile, recent advances in image-based neural rendering have achieved impressive success in hair view synthesis. By employing primitives specifically tailored for strand structures, such as hybrid neural volumes~\cite{wang2022hvh}, neural strand features~\cite{rosu2022neural}, and linked cylindrical Gaussians~\cite{luo2024gaussianhair}, these methods exhibit a superior capability in capturing fine-grained appearance details from multi-view inputs. However, lacking a physics-grounded appearance parameterization, these primitives tend to entangle geometry, appearance, and associated illumination conditions, leading to overfitting of input images at the expense of generalization capability. Consequently, significant quality degradation is constantly observed in their relighting results.

Towards high-quality hair relighting, our motivation is to leverage the complementary strengths of both physically based material models and neural rendering representations. This involves employing analytical BSDF models to reconstruct the primary light transport while relying on neural rendering to capture subtle appearance details.

In this work, we introduce \textit{GroomLight}, a hair appearance modeling method that reconstructs relightable hair appearance from OLAT (one-light-at-a-time) images. For the very first time, \textit{GroomLight} enables high-fidelity and fully functional relighting for photorealistic human hair. Our appearance representation consists of two key components: 1) an \textit{extended hair BSDF model} robust to natural hair color variations and imperfections in the reference hair geometry; and 2) a \textit{light-aware residual model} that captures remaining details utilizing light- and view-conditioned dual-level spherical harmonics (SH) associated with a Gaussian-based hair representation. On top of that, we propose a hybrid inverse rendering pipeline that combines a physics-based differentiable renderer with neural image-based rendering to optimize the appearance.

We extensively evaluate our method on real-world hair data, demonstrating state-of-the-art relighting and view synthesis results under diverse lighting conditions. Moreover, our pipeline naturally supports versatile applications, including material editing and dynamic animation rendering, and seamlessly integrates with conventional CG workflows.

%% file: sec/2_related.tex
\section{Related Work}
\label{sec:related work}

\subsection{Hair Appearance Modeling}
Besides geometry capture from single-~\cite{chai2012single,chai2013dynamic,hu2015single,chai2016autohair} or multi-view~\cite{paris2008hair,jakob2009capturing,luo2013structure,zhang2017data,nam2019strand}, hair appearance modeling has also been a long-standing challenge.
We provide a brief overview of physics- and neural-based methods that model the complex appearance of hair fibers.

\noindent\textbf{Physically Based Models.} Physically based hair appearance models describe light transport resulting from scattering on and within hair fibers. Building upon the foundational work on fiber reflectance~\cite{kajiya1989rendering}, \citet{marschner2003light} and subsequent research~\cite{chiang2015practical, d2011energy, d2014fiber, huang2022microfacet, sadeghi2010artist, zinke2007light, zinke2009practical, khungurn2017azimuthal, zhu2022practical} have derived hair BSDFs that encapsulate hair scattering effects into analytical reflection and transmission lobes. These BSDF models, when used in conjunction with a path tracer~\cite{Kajiya1986}, can generate highly realistic images of hair. Variations of hair BSDF models have also been employed to render textile fibers~\cite{aliaga2017appearance} and animal fur~\cite{yan2015physically, yan2017efficient}. While most models rely on geometric optics, incorporating wave-optical effects can further enhance realism~\cite{xia2020wave, xia2023practical, benamira2021combined}. Nevertheless, analytical physically based BSDFs may lack the expressive capacity and flexibility to perfectly match the appearance of real hair, potentially leading to inaccuracies in re-rendering crucial appearance features such as specular highlights and color variations.

\noindent\textbf{Neural Representations.} Neural networks offer a powerful means of representing high-fidelity human hair appearance. Generative models~\cite{chai2020neural, wei2018real, jo2019sc, tan2020michigan, qiu2019two} have been trained for realistic hair synthesis. Neural rendering techniques~\cite{tewari2022advances} leverage scene representations such as neural radiance field (NeRF)~\cite{mildenhall2020nerf, rosu2022neural, ma2021pixel}, 3D Gaussian Splatting (3DGS)~\cite{kerbl3Dgaussians, xu2024gaussian, luo2024gaussianhair, qian2024gaussianavatars}, and implicit geometry and texture networks~\cite{zheng2022avatar, grassal2022neural, chan2022efficient, sun2023next3d} to learn portrait appearance from images. While neural models can produce high-quality, photorealistic results, they often lack robust relighting capabilities and can be challenging to animate or edit manually.

\input{figs/fig_pipeline}
\subsection{Inverse Rendering of Hair}
Inverse rendering aims to estimate appearance parameters from visual observations using \emph{analysis-by-synthesis} techniques. \citet{sun2021human} estimate hair BSDF~\cite{chiang2015practical} parameters by measuring differences between sparse hair samples and synthetic ground truth observations. Their approach relies on an approximation of the observed color based on idealized lighting conditions and demonstrates results on synthetic data with accurate hair geometry reconstructed from ground truth segmentation, camera and lighting information. \citet{chai2015high} propose an image-based rendering technique for hair relighting. Neural-based relighting~\cite{zhang2021physg, sun2021nelf, munkberg2022extracting, bi2020neural, zhang2021nerfactor} and Gaussian-based inverse rendering methods~\cite{saito2024relightable, bi2024rgs, shi2023gir, jiang2024gaussianshader, gao2023relightable, zheng2024physavatar} often combine neural representations with simplified physically based rendering functions, which may not accurately model complex phenomena such as multiple scattering and shadowing. \citet{luo2024gaussianhair} leverage 3D Gaussians to capture hair appearance and design a scattering approximation method for relighting. However, their method remains fundamentally based on Gaussian splatting with baked-in appearance, and the approximation function may not fully capture the intricacies of real-world light scattering in human hair, potentially leading to suboptimal rendering results. In contrast, we propose a hybrid inverse rendering pipeline that fully leverages the strengths of both physically based and neural rendering representations.

%% file: figs/fig_pipeline.tex
\begin{figure*}[ht!]
    \centering
    \includegraphics[width=0.94\linewidth]{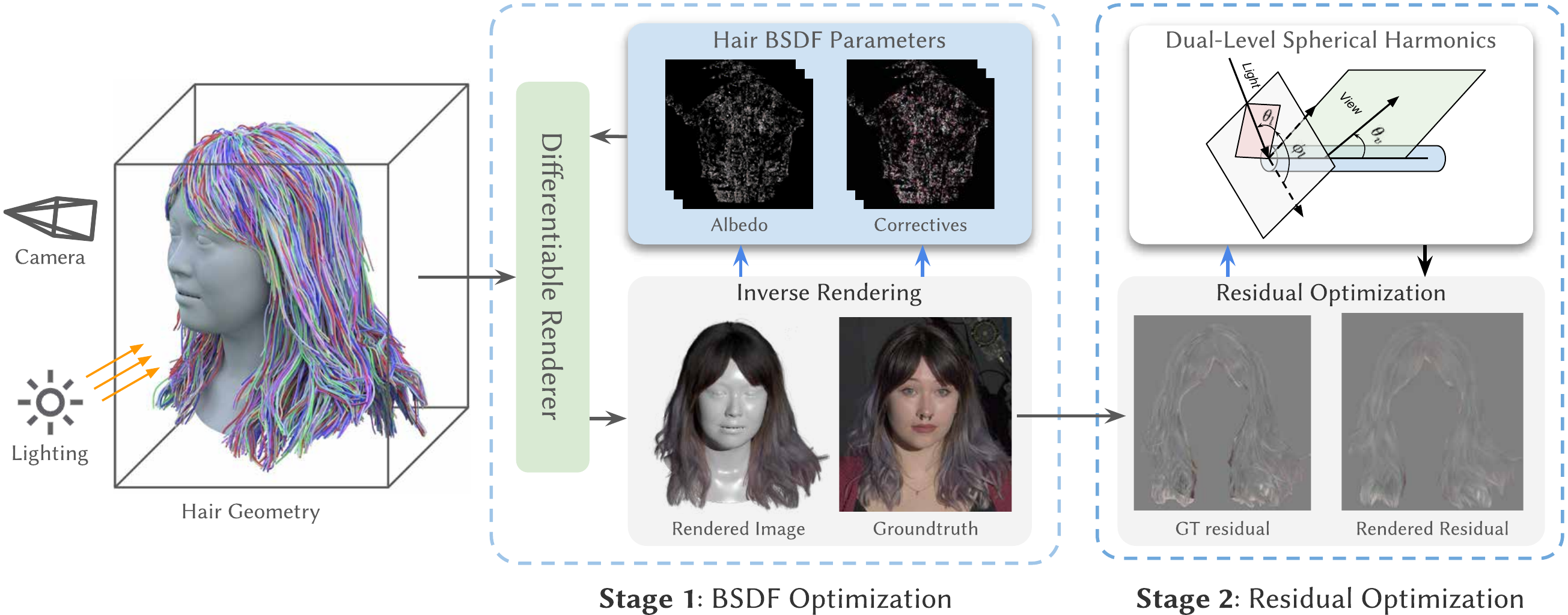}
    \caption{\textbf{Our hybrid inverse rendering pipeline} employs a two-stage optimization scheme. Given input OLAT images and the reconstructed hair geometry (here we visualize the 1/10 downsampled version), we first estimate the physical parameters of the extended hair BSDF model (Stage 1, \S\ref{Sec3.2}), and then leverage a light-aware residual model to capture fine-grained appearance details (Stage 2, \S\ref{Sec3.3}).}
    \label{fig:overview}
\end{figure*}

%% file: sec/3_method.tex
\section{Method}

\subsection{Overview}
\label{Sec3.1}
Our approach employs a two-level representation of hair appearance. This comprises: 1) an extended BSDF model (\S\ref{Sec3.2}), which utilizes a physics-based parameterization to capture the primary appearance characteristics of real-world hair strands; and 2) an implicit light-aware residual model (\S\ref{Sec3.3}) that captures light- and view-dependent details not fully encompassed by the analytical BSDF formulation.

Accordingly, our hybrid inverse rendering pipeline (\S\ref{Sec3.4}) constructs these appearance models in two stages. First, inverse rendering optimization is performed within the analytical BSDF parameter space, which is extended to accommodate spatial variations in hair material and improve robustness against potential misalignment in the input hair geometry. Subsequently, leveraging a 3D Gaussian-based representation of individual strands, we optimize light-aware dual-level spherical harmonics to model the remaining appearance residuals.

\input{figs/fig_scattering}
\paragraph{Input Data.} Our method operates on multi-camera OLAT captures. We capture our input data in a light stage, featuring an array of 50 cameras and 331 light sources, uniformly distributed over the dome. From the complete set of frames $\mathcal{I}$, we exclude views where cameras do not prominently feature hair or where the subject is strictly back-lit (to mitigate artifacts arising from low exposure). The reference hair geometry $\mathcal{S}$ and parametric head mesh $\mathcal{H}$ are pre-reconstructed from the same OLAT input using the state-of-the-art strand-level hair capture technique of \textit{GroomCap}~\cite{zhou2024groomcap}, which typically yields 150K scalp-rooted strands. Additionally, we perform 2D segmentation using \textit{SAM}~\cite{kirillov2023segment} to obtain a set of hair masks $\mathcal{M}$ for input frame images.

\subsection{Extended Hair BSDF Model}
\label{Sec3.2}
Our strand-based hair BSDF is based on the analytical fiber reflectance model by~\citet{chiang2015practical}. Each strand's geometry is approximated as a cylindrical B-spline curve with a uniform radius, which we set to the average radius of human hair (40$\mu m$). For each point on the fiber, the perceived reflectance is parameterized as a function of the incident $\theta_i$ and exitant $\theta_o$ lighting inclinations, their relative azimuth $\phi = \phi_o - \phi_i$, and the physical parameters $\Psi$ of the material:
\begin{equation}
    \text{S}(\theta_i,\theta_o,\phi;\Psi)=\sum_{p=0}^\infty \text{M}_p(\theta_i,\theta_o;\Psi)\text{N}_p(\phi;\Psi),
    \label{eq:bsdf}
\end{equation}
where the reflectance $S$ is decomposed into separate lobes according to the path of light propagation $p$ (i.e., $p\in\{R, TT, TRT, \dots\}$), which counts the reflections and transmissions in the fiber cylinder. Each lobe consists of $\text{M}_p$ and $\text{N}_p$, i.e., the longitudinal and azimuthal scattering functions, respectively. \cref{fig:scattering} shows the light scattering effects in hair rendering, where single-bounce illumination (only $p=0$ is evaluated) leads to degenerated results, while more scattering events contribute to more realistic rendering. 

The material parameters $\Psi$ of the BSDF model, to be optimized during inverse rendering, include absorption coefficients $\sigma_a$, longitudinal $\beta_M$ and azimuthal $\beta_N$ roughnesses, and interior $i_{ior}$ and exterior $e_{ior}$ indices of refraction~\cite{chiang2015practical}.

\paragraph{Scalp-Aligned Albedo Parameterization.}
By assuming homogeneous material parameters $\Psi$ (Eq.\ref{eq:bsdf}), a uniform hair albedo is enforced across all strand points, which cannot model the nuanced color variations commonly observed in real world, naturally or artificially, ranging from root-to-tip gradients within a single strand to subtle differences across wisps. To address this limitation, following previous work~\cite{wang2009example, zhou2018hairnet, rosu2022neural, zhou2024groomcap}, we propose a scalp-aligned parameterization that introduces sufficient degrees of freedom to capture spatial color variation, while maintaining natural regularization and preventing over-parameterization.

Specifically, recall that all input strands $\mathcal{S}$ are rooted on the parametric head $\mathcal{H}$. We pre-define a scalp region $\mathcal{H}_S$ on the head surface and unwrap it to a 2D parameterization, namely the \textit{scalp space}. For each strand $s\in \mathcal{S}$, we project its root to the corresponding attachment point on $\mathcal{H}_S$ and embed it into the scalp space at coordinates $u_s$ via barycentric interpolation.
We also convert each strand $s$ to a B-spline curve $s^*$ by uniformly sampling $n_c$ control points.
The hair is finally formulated as a regular 3D volume, conceptually equivalent to $n_c$ layers of textures with resolutions of $(w_u, h_u)$. An arbitrary point $p$ on strand $s$ is mapped to this volume based on the strand's root coordinate $u_s$ and the point's relative on-strand coordinate $w_p^s$. We utilize this scalp-aligned parameterization to encode absorption coefficients $\boldsymbol{\sigma}_a\in \mathbb{R}^{n_c\times w_u\times h_u}$, which determine the albedo.

Compared to world-space voxelization, this parameterization ensures that points on the same strand share the same texture coordinate, and nearby texels map to points on adjacent strands. Decoupled from specific geometric styles, this inherent structure leads to more controllable degree of freedom and effective regularization across texture layers.

\paragraph{Rotation Correctives.}
In contrast to previous works \cite{sun2021human} that benefit from clean images and relatively precise hair geometry obtained from synthetic renderings, our pipeline must contend with imperfections in hair reconstruction that introduces discrepancies between the reconstructed and ground-truth strand shapes. This poses challenges for inverse rendering, particularly problematic for specular highlights, which are highly sensitive to the accuracy of strand tangents. To mitigate sub-optimal roughness optimization due to misplaced highlights, we propose a lightweight corrective to the BSDF model that, rather than directly refining the curve geometry, optimizes a constrained rotation transformation for each strand segment to compensate for inaccuracies in the tangent directions.

Specifically, we formulate this rotation corrective $\boldsymbol{R}$ within the same scalp-aligned parameterization used for the absorption coefficients $\boldsymbol{\sigma}_a$. This corrective  encodes a rotation matrix $R$ for each strand control point, represented using the Rodrigues formula. When evaluating a BSDF sample, its local shading frame $\{n,t,b\}$, defined by the normal $n$, tangent $t$, and bitangent $b$, is rotated by $R$ to yield a corrected frame $\{Rn,Rt,Rb\}$ for shading.

\subsection{Light-Aware Residual Model}
\label{Sec3.3}
While our BSDF model effectively reconstruct the large-scale appearance of hair and generalizes well to novel views and lighting conditions, relying solely on this analytical model leads to over-smoothed results due to a broad range of imperfect assumptions such as simulated lighting, missing diffraction and interferenc in the BSDF model, etc., which limits its capacity to capture subtle, yet visually critical, local details. 

We introduce a residual term to model the remaining discrepancies between the physics-based rendering obtained the optimized BSDF and the real-world observations. To ensure desired consistency and generalization, this residual model is designed to be both light-aware and strand-aligned, to enable robust relighting capabilities and facilitate a compact, geometry-aware parameterization.

Specifically, we leverage 3D Gaussian Splatting (3DGS) \cite{kerbl3Dgaussians} for scene representation. However, instead of optimizing free-form Gaussians as in the vanilla 3DGS formulation, we enforce identical anisotropic Gaussians for all kernels, and place them along the input strand geometry to form a chained sequence. During optimization, we fix the geometric parameters of all Gaussians (position, covariance, and opacity) and focus solely on optimizing their appearance parameters, as detailed below. This strategy is motivated by our primary focus on appearance modeling, recognizing that the initial Gaussian configuration already inherits good strand geometry from the input.

\paragraph{Dual-Level Spherical Harmonics.}
We introduce a dual-level spherical harmonics (SH) function as the appearance representation for the residual model, which is both view- and light-dependent, achieved by nesting two levels of SH. Given a dual-level SH, denoted as $\text{C}_{vl}$, the final Gaussian color $c$ is calculated as a function of the camera view $d_v$ and incident lighting $d_l$ directions (in global coordinates):
\begin{equation}
\begin{split}
    c=\text{C}_v(d_v),\\
    \text{C}_v=\text{C}_{vl}(d_l).
\end{split}
\label{eq:shs}
\end{equation}

With both SH levels using the same degree $l$, the parameter size of each dual-level SH is $\text{C}_{vl}\in\mathbb{R}^{(l+1)^2\times(l+1)^2\times3}$ and $\text{C}_{v}\in\mathbb{R}^{(l+1)^2\times3}$.
To maintain compact appearance parameters, we attach $k$ dual-level SHs to each strand as anchor SHs.
During rasterization, for each Gaussian primitive at location $p$ belonging to strand $s$, its dual-level SH parameters are linearly interpolated from its nearest two anchor SHs based on its relative on-strand coordinate $w^s_p$.

\subsection{Hybrid Inverse Rendering}
\label{Sec3.4}
Given the input hair geometry $\mathcal{S}$ and multi-camera OLAT images $\mathcal{I}$, we optimize the hair appearance, including both the analytical BSDF model and the residual model, using our hybrid inverse rendering in two stages.

\paragraph{Stage 1: BSDF Optimization.}
The first stage optimizes the physical parameters $\Psi$ of the BSDF model using differentiable path tracing. Due to the large number of scattering events, we use path replay backpropagation~\cite{Vicini2021PathReplay} to efficiently compute gradients of the used photometric loss:
\begin{equation}
    \mathcal{L}_{rgb}=\sum_{I\in\mathcal{I}}\left\|I-\text{R}(\Psi,\mathcal{S},\mathcal{C}(I),\mathcal{L}(I)) \right\|_1,
    \label{eq:l1 loss}
\end{equation}
where $\text{R}$ represents the rendering function, $I\in\mathcal{I}$ denotes one ground truth view image, and $\mathcal{C}(I)$ and $\mathcal{L}(I)$ are the calibrated camera and lighting information associated with the image $I$, respectively.

With our albedo parameterization and rotation correctives, the complete set of parameters for the hair BSDF model is represented as $\Psi=\{\boldsymbol{\sigma},\boldsymbol{R},\beta_N,\beta_M,e_{ior},i_{ior}\}$. To ensure smooth variation along each strand, we add regularization terms for the spatially varying variables in the scalp-aligned volume:
\begin{equation}
    \mathcal{L}_{\sigma} = \sum_{i=1}^{n_c-1}\left\| \boldsymbol{\sigma}_i - \boldsymbol{\sigma}_{i-1}\right\|_1,
\end{equation}
\begin{equation}
    \mathcal{L}_{R} = \sum_{i=1}^{n_c-1}\left\| \boldsymbol{R}_i - \boldsymbol{R}_{i-1}\right\|_1.
    \label{eq:reg loss}
\end{equation}

Finally, the complete loss function for the first inverse rendering stage is:
\begin{equation}
    \mathcal{L}_{BSDF}=\mathcal{L}_{rgb}+\lambda_{\sigma}\mathcal{L}_{\sigma}+\lambda_{R}\mathcal{L}_{R},
    \label{eq:full loss}
\end{equation}
where $\lambda_{\sigma} = 0.5$ and $\lambda_{R} = 10$ control the strength of both regularization terms.

\paragraph{Stage 2: Residual Optimization.} Once the BSDF optimization in Stage 1 is complete, we proceed to optimize the residual model. This involves fitting the remaining differences between the rendered images and the input views. Specifically, we optimize the dual-level spherical harmonics parameters of the residual appearance by minimizing the following residual loss:
\begin{equation}
    \mathcal{L}_{res}=\sum_{I\in\mathcal{I}}\left\|I-(I_{render} + I_{res})\right\|_2^2,
    \label{eq:l1 loss}
\end{equation}
where $I_{render}$ represents the path-traced rendering using the optimized BSDF parameters from Stage 1, and $I_{res}$ is the residual image rendered via 3D Gaussian rasterization. Note that in stage 2 we employ $L_2$ loss which we find leads to more robust performance.

\paragraph{Implementation Details.} The inverse rendering pipeline in Stage 1 is built upon the differentiable renderer of Mitsuba 3~\cite{Mitsuba3}. We approximate the lighting using measured IES profiles, with lighting positions initialized via OLAT calibration. We jointly optimize the BSDF parameters $\Psi$ and the lighting parameters, including positions, orientations, and intensities since the lighting calibration might not be perfectly accurate. In each optimization iteration, we randomly select 8 images from all camera views and lighting conditions, and aggregate gradients across these images. For a given capture, we run 1000 iterations, which takes about 20 hours on a single NVIDIA A100 GPU (40 GiB).

In Stage 2, we downsample the input hair to approximately 52K strands, each represented by 100 segments. Cylindrical Gaussians are attached to these strand segments, with their positions and covariance matrices derived from the location and orientation of the the corresponding hair segments. Each strand is associated with 8 anchor SHs with SH degree $l=3$, and per-segment SH parameters are obtained through interpolation between its two nearest anchor SHs. During optimization, we optimize the SH appearance parameters and keep all geometric properties fixed. We train the residual Gaussians for 160K iterations on 8 NVIDIA A100 GPUs (40 GiB), which takes about 12 hours. 

%% file: figs/fig_scattering.tex
\begin{figure}[t!]
    \centering
    \includegraphics[width=\linewidth]{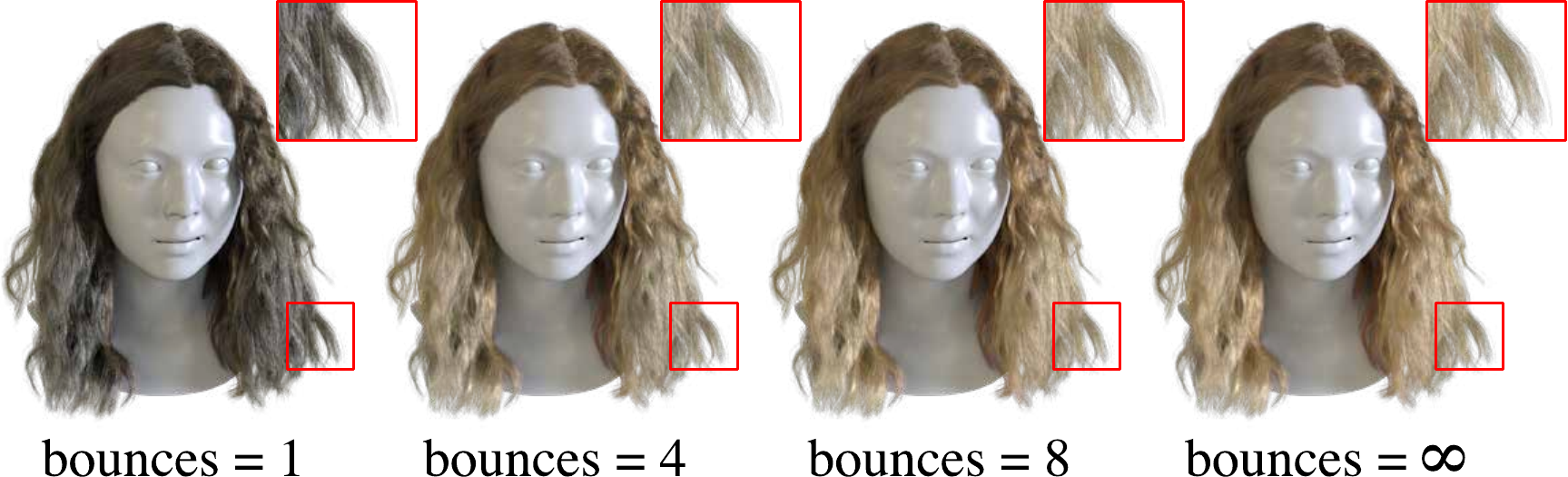}
    \caption{\textbf{Impact of light scattering.} We compare the effects of varying the maximum number of ray bounces during path tracing ($\infty$ means the integrator will not enforce any hard cutoff on the number of bounces). Our method employs 8 bounces to achieve high quality rendering at practical efficiency.} 
    \label{fig:scattering}
\end{figure}

%% file: sec/4_experiments.tex
\section{Experiments}
In this section, we first explain our experimental setup, including the dataset, baselines, and evaluation metrics. Then we present and discuss the experiment results, including comparisons and ablation study, in detail.

\input{figs/fig_results}
\input{tabs/results}
\input{figs/fig_ablation}

\subsection{Setup}
\paragraph{Dataset.} We capture our test data from 10 subjects (\S\ref{Sec3.1}). These subjects feature diverse hair characteristics, including variations in length (short, medium, and long), style (curly, wavy, and straight), and color (black, blond, and gray). The hair geometry is reconstructed by \textit{GroomCap}~\cite{zhou2024groomcap} from views under uniform lighting. For each subject, we use 160 lighting conditions and approximately 48 cameras views for training. We designate 2 novel views (one front and one back view) and 160 novel lightings for evaluation. 

\paragraph{Baselines.} To assess the quality of view synthesis and relighting, we compare \textit{GroomLight} with two adapted state-of-the-art baselines quantitatively and qualitatively:
\begin{itemize}
    \item \textit{HairInverse}~\cite{sun2021human}: An inverse rendering method based on an analytical BSDF material model. The original method performs image-space optimization on selected pixels without involving differentiable renderers and presents results only on synthetic data. We re-implement this baseline, adapting to our algorithm, which iterative optimizes BSDF parameters based on photometric losses.
    \item \textit{GaussianHair}~\cite{luo2024gaussianhair}: A 3DGS-based~\cite{kerbl3Dgaussians} neural rendering method for hair. We re-implement this baseline by reusing our chained cylindrical Gaussians and applying our residual appearance model, i.e., dual-level SH, to the full image. This differs from their hair scattering model, which employs optimized SH base colors and manually tuned material parameters.
\end{itemize}

\input{figs/fig_results_edit}
\paragraph{Evaluation Metrics.} We evaluate all methods using the following photometric error metrics: Peak Signal-to-Noise Ratio (PSNR), Structural Similarity Index Measure (SSIM), and Learned Perceptual Image Patch Similarity (LPIPS).

\subsection{Comparisons}
We compare our method with both \textit{HairInverse} and \textit{GaussianHair} baselines on testing view reconstruction under diverse lightings.
As demonstrated in \cref{tab:quantitative}, our method outperforms both baselines by a large margin across all metrics.
We also show qualitative comparisons in \cref{fig:results}.

\textit{HairInverse}~\cite{sun2021human}, based on homogeneous hair BSDF material, estimates its parameters from the average color of the hair region. This model lacks the flexibility to represent the diverse textures found in real-world hair.

While 3DGS-based \textit{GaussianHair}~\cite{luo2024gaussianhair} achieves reasonable view reconstruction results, it often overfits to training views and lightings, leading to poor generalization.

As shown in \cref{fig:results}, \textit{GroomLight} effectively reconstructs coarse hair appearances in stage 1 (\textit{Ours Stage1}). By incorporating the residual model in Stage 2, we achieve photorealistic final renderings (\textit{Ours Full}) with enhanced details.

\subsection{Ablation Study}
We conduct an ablation study to evaluate the contribution of each proposed component in our method, including the spatially varying albedo parameterization and rotation correctives in Stage 1, and the residual module of Stage 2.

\paragraph{Extended BSDF.} We augment the base hair BSDF with a scalp-aligned albedo parameterization to enable spatially-varying hair color and introduce rotation correctives to compensate for inaccuracies in the reference geometry. We present the performance of different variants in \cref{tab:quantitative}: including \textit{Stage1 OrigBSDF} (Stage 1 only using the original version of hair BSDF~\cite{chiang2015practical}), \textit{Stage1 NoCorr} (adding the albedo parameterization), \textit{Stage1 Full} (adding both the albedo parameterization and rotation correctives). The results demonstrate that each component contributes to improved reconstruction quality. As shown in \cref{fig:results_ablation}, the original hair BSDF, assuming homogeneous materials, produces a uniform color across all strands. The spatially-varying albedo facilitates the reconstruction of more detailed hair texture, while the rotation correctives further enhance the accuracy of light reflectance, particularly for highlights. 

\input{figs/fig_sim}

\paragraph{Residual Model.} We incorporate the residual model to address the inherent limitations of analytical BSDF models, which are based on approximations of physical behaviors. This residual model helps capture fine-grained texture details and highlights (\textit{Full}), leading to more accurate results (\cref{tab:quantitative}) and enhanced photorealism as shown in \cref{fig:results_ablation}.

\subsection{Applications}
The generalization capabilities of our hair appearance model naturally lends itself to various downstream applications that are previously challenging.

\paragraph{Relighting.} \cref{fig:teaser} shows our relighting results under diverse environment lighting conditions. To provide the lighting direction input for our residual module, we approximate the light source in HDR environment maps as the position with maximum radiance. Our method achieves high-fidelity rendering quality across a wide range of lighting scenarios.

\paragraph{Appearance Editing.} Our spatially-varying albedo parameterization enables convenient appearance editing (\cref{fig:results_edit}). By directly modifying the absorption coefficients for hair strands on the scalp space embedding, we can achieve various color effects. In addition to albedo editing, our representation supports rendering diverse highlight patterns by changing the roughness parameters.

\paragraph{Dynamic Rendering.} Our method also generalizes to novel hair geometry. By animating the reconstructed hair using off-the-shelf simulation tools, we can render the sequence using the optimized appearance model, including BSDF and residual parameters, as shown in \cref{fig:resultssim}. Please refer to the supplementary video for the entire sequence.

%% file: figs/fig_results.tex
\begin{figure*}[h!]
    \centering
    \includegraphics[width=0.97\linewidth]{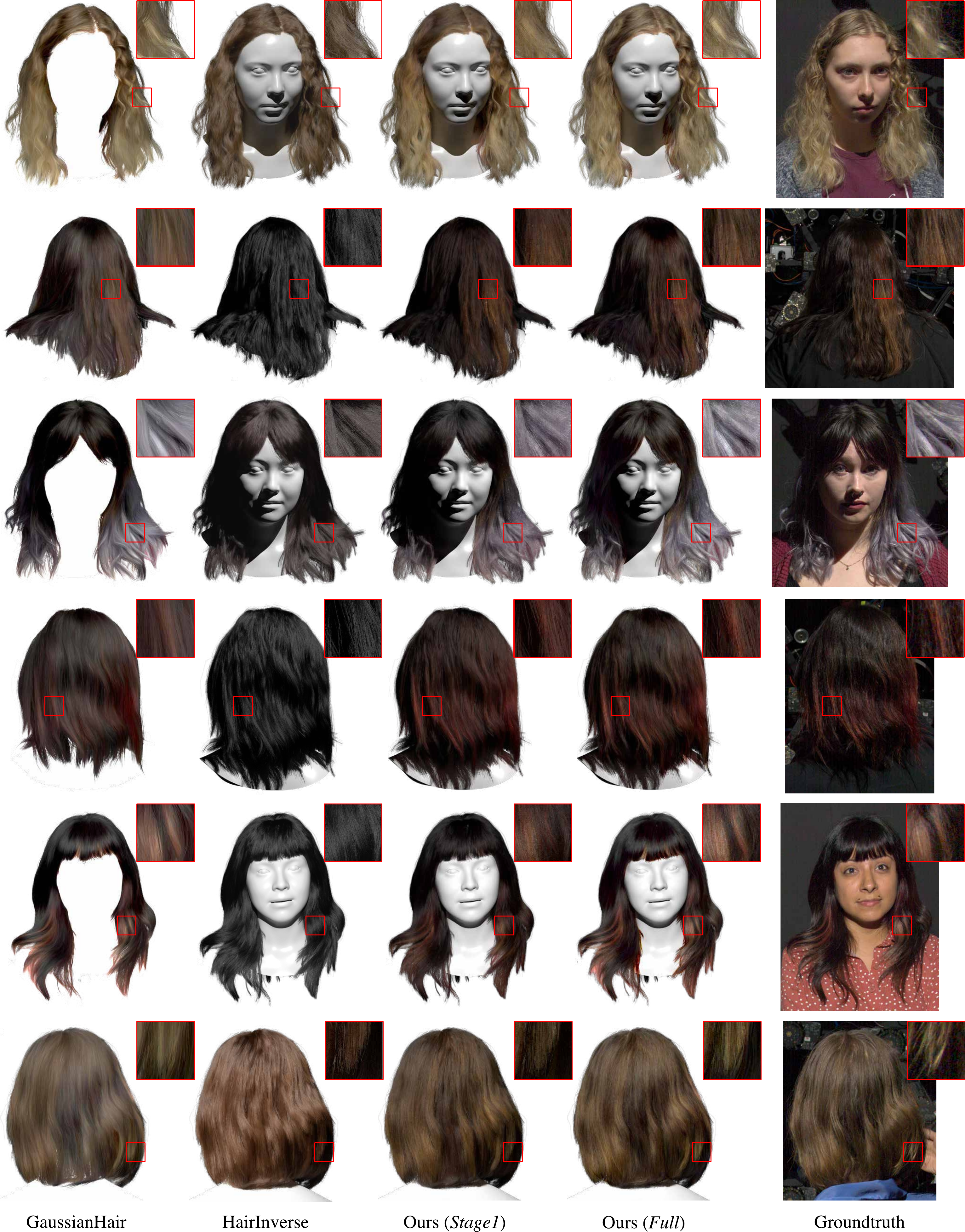}
    \caption{\textbf{Qualitative evaluation} of rendering results at novel testing views under novel lighting conditions of our method and baselines.} 
    \label{fig:results}
\end{figure*}

%% file: tabs/results.tex
\begin{table}[t!]
    \centering
    \resizebox{0.5\textwidth}{!}{
    \begin{tabular}{cccc}
    \toprule
    Method & PSNR (\(\uparrow\))  & SSIM (\(\uparrow\)) & LPIPS (\(\downarrow\)) \\
    \midrule
    HairInverse~\cite{sun2021human} & 26.50	& 0.6767 & 0.3171\\
    GaussianHair~\cite{luo2024gaussianhair} & 27.22 & 0.7340 & 0.2920 \\
    \midrule 
    Ours (\textit{Stage1 OrigBSDF}) & 27.75 & 0.7350 & 0.2772 \\	
    Ours (\textit{Stage1 NoCorr}) & 28.58 & 0.7410 & 0.2690 \\
    Ours (\textit{Stage1 Full}) & 28.67 & 0.7423 & 0.2653 \\ 
    \midrule 
    Ours (\textit{Full}) &  \textbf{29.71} & \textbf{0.7727} & \textbf{0.2461} \\ 
    \bottomrule
    \end{tabular}
    }
    \caption{\textbf{Quantitative results on view reconstruction} of baseline comparisons and ablation study.} 
    \label{tab:quantitative}
\end{table}

%% file: figs/fig_ablation.tex
\begin{figure*}[t!]
    \centering
    \includegraphics[width=\linewidth]{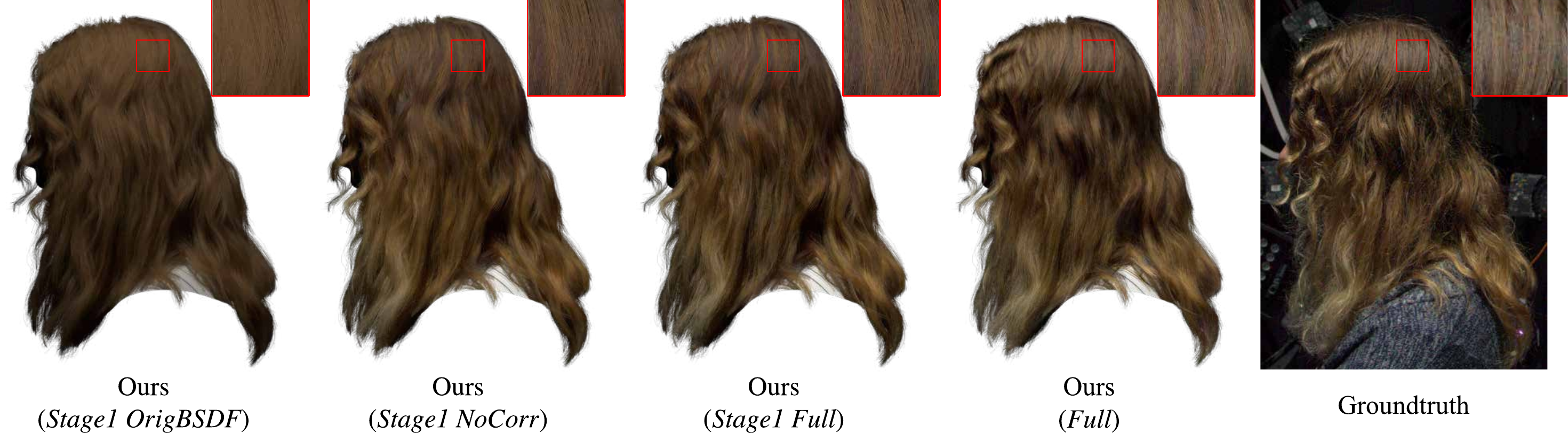}
    \caption{\textbf{Ablation study} of components in our method. Starting from Stage 1 only with original BSDF model (\textit{Stage1 origBSDF}), we sequentially add spatially-varying albedo (\textit{Stage1 NoCorr}), rotation correctives (\textit{Stage1 Full}), and the residual model (\textit{Full}).}
    \label{fig:results_ablation}
\end{figure*}

%% file: figs/fig_results_edit.tex
\begin{figure}[t!]
    \centering
    \includegraphics[width=\linewidth]{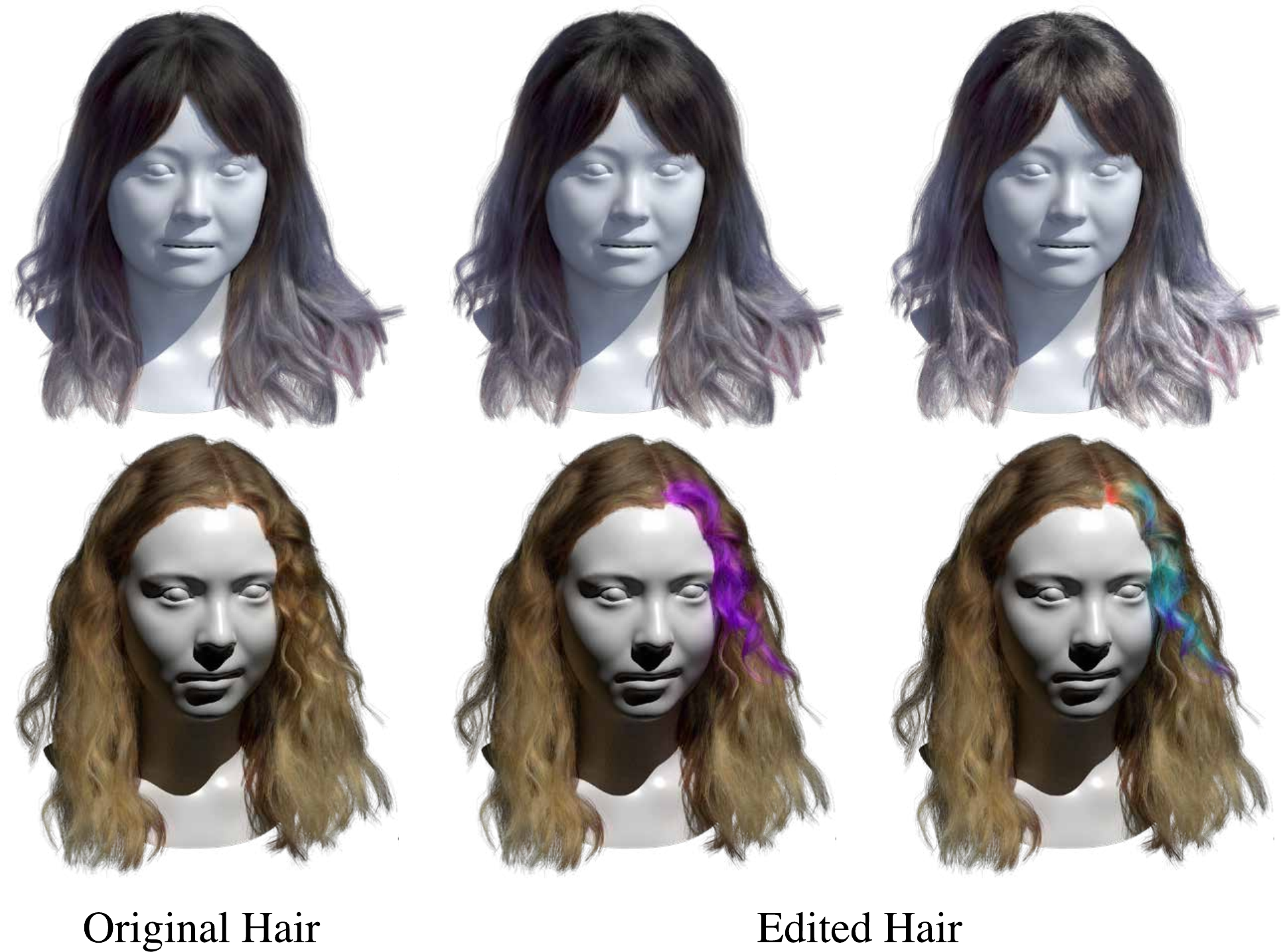}
    \caption{\textbf{Appearance editing}. Top row: decreasing roughness of hair BSDF to add more highlights. Bottom row: editing the albedo parameters to change the color for specific hair wisps.}
    \label{fig:results_edit}
\end{figure}

%% file: figs/fig_sim.tex
\begin{figure}[t!]
    \centering
    \includegraphics[width=\linewidth]{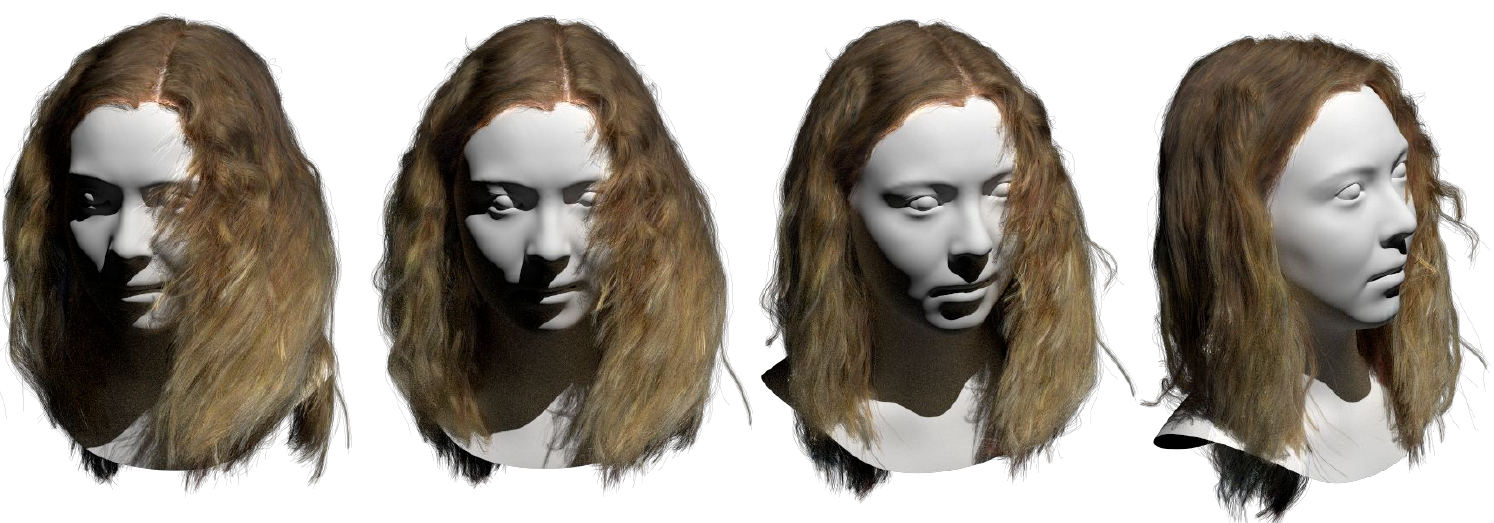}
    \caption{\textbf{Dynamic rendering}. We show rendering results of animated hair sequences using our optimized appearance model.}
    \label{fig:resultssim}
\end{figure}

%% file: sec/5_conclusion.tex
\section{Conclusion}

We introduce \textit{GroomLight}, a novel method for capturing and relighting photorealistic hair. Our approach combines the strengths of both physically based rendering and neural representation learning within a hybrid inverse rendering framework. We propose an extended analytical BSDF model that effectively captures the primary appearance of hair while remaining robust to variations in hair color and inaccuracies in the underlying geometry. To further enhance visual fidelity, we introduce a light-aware residual model based on dual-level spherical harmonics and a 3D Gaussian-based scene representation.
\textit{GroomLight} achieves state-of-the-art performance on inverse rendering for relightable human hair appearance modeling from real-world input data, which could benefit a wide specturm of applications in virtual reality, digital fashion, movie industry, and more.

\paragraph{Limitations and Future Work.}
\textit{GroomLight} relies on light stage capture setup to obtain lighting and camera calibrations, which limits its applicability to in-the-wild data where lighting and camera information are unknown. In addition, our pipeline is based on offline optimization and renders images at about 1 FPS during inference, which cannot support real-time applications. Furthermore, our method requires relatively accurate geometry and segmentation masks to prevent artifacts arising from the inclusion of unwanted textures from clothing, the face, or the background. We aim as future work to improve the efficiency of our pipeline, leverage large-scale datasets and neural networks to enable generalization of our method for in-the-wild image inputs. 

%% file: sec/6_ack.tex
\section*{Acknowledgement}

We thank Tiancheng Sun, Stefanos Zafeiriou, and Erroll Wood for their helpful discussions; Jay Busch, Cynthia Herrera, Christoph Rhemann, Di Qiu, and Xu Chen for their assistance with data capture and processing; and all of our capture models. We are also grateful for the insightful feedback from the anonymous reviewers.

%% file: sec/suppl.tex
\section{Implementation Details}
In this section, we provide additional implementation details about our method and the baselines.

\subsection{Our Method} 
Our inverse rendering pipeline consists of a two-stage optimization scheme. In \textbf{\textit{Stage 1}}, we estimate the physical parameters of the extended hair BSDF model (Sec 3.2). We build upon the original hair BSDF~\cite{chiang2015practical}, enabling albedo parameterization to support diverse hair colors and introducing rotation correctives to compensate for imperfect geometry input. Specifically, the albedo parameterization is implemented as a texture map, and rotation correctives are represented by a 2D rotation map similar to a normal map in common BRDF models. Both maps are defined in $4096\times4096$ resolution. When evaluating a shading frame for the 3D point on the hair B-spline curves that intersects with the ray, we first find its rotation corrective using the rotation map and the point's UV coordinates. We then apply the rotation to the shading frame to change the incident and exitant lighting directions, which are used for evaluating the hair BSDF. The entire process is differentiable, as implemented in Mitsuba 3~\cite{Mitsuba3} based on path replay backpropagation~\cite{Vicini2021PathReplay}. Given multi-view OLAT images, we optimize the hair BSDF parameters using the L1 photometric loss and regularization losses (Sec 3.4). Note that we jointly optimize the lighting parameters, which we find leads to better performance. The optimization is performed on a single NVIDIA A100 GPU (40\,GiB).

In \textbf{\textit{Stage 2}}, we introduce a residual model based on the 3D Gaussian representation~\cite{kerbl3Dgaussians} with dual-level spherical harmonics to enhance rendering photorealism. We utilize eight NVIDIA A100 GPUs (40\,GiB) for training and evaluation. During training, we randomly split the hair strands and equally distribute them across six A100 GPUs. We use the remaining two GPUs for splatting and computing gradients, respectively. In each training iteration, we randomly choose one split and render the residual map to compute the L2 photometric loss. Note that we employ the L2 loss instead of L1, as we find that the L1 loss leads to overly noisy results. During evaluation, we combine all the splits for rendering.

\subsection{Baselines} We compare our method to current state-of-the-art methods, including HairInverse~\cite{sun2021human} and GaussianHair~\cite{luo2024gaussianhair}. Since no public code is available, we re-implement the methods based on the algorithms described in the papers.

\vspace{3mm}\noindent\textbf{HairInverse}~\cite{sun2021human} is the inverse rendering pipeline introduced in HairInverse consists of geometry reconstruction based on a synthetic setup and estimation of hair BSDF~\cite{chiang2015practical} parameters. However, the method only estimates the average color of the hair and struggles with real-world data due to failures in geometry reconstruction. To ensure fair comparisons, we use our reconstructed hair geometry as input and employ our optimized lighting and hair roughness parameters to render results for evaluation. The optimization process is performed using views under uniform lighting, following Algorithm 1 from the paper~\cite{sun2021human}. We build the training and testing pipelines in Mitsuba 3~\cite{Mitsuba3}.

\vspace{3mm}\noindent\textbf{GaussianHair}~\cite{luo2024gaussianhair} includes geometry and appearance reconstruction based on connected 3D Gaussians~\cite{kerbl3Dgaussians}, similar to our chained cylinder Gaussian representation. To enable relighting, the method introduces a hair scattering function that employs optimized SH base colors and manually tuned material parameters. For simplicity, we leverage our Gaussian representation with dual-level spherical harmonics to evaluate the performance of relightable 3D Gaussians in modeling human hair appearance. We use the same inputs as our method (hair geometry and multi-view OLAT images) to train the model. Similarly, we solely optimize the spherical harmonics with other parameters (i.e., 3D Gaussian positions, covariance matrices and opacities) fixed.

\section{Additional Results}
We highly recommend watching the supplementary video for more results on relighting, appearance editing, dynamic rendering, and evaluations.

We present more qualitative comparison results with baselines in \cref{fig:results_relighting}. In addition, we visualize the error maps of view synthesis evaluation, for both baseline comparisons (\cref{fig:error_baseline}) and ablation study (\cref{fig:error_ablation}). Consistently, our method achieves the best results both visually and quantitatively, against all baselines we have evaluated.

We also include additional visual results. In \cref{fig:results_relighting}, we show relighting results under diverse environmental lighting and rotated hair geometry. Our results demonstrate high-fidelity rendering under novel lighting conditions and novel view configurations. In \cref{fig:results supp 1}, we present more qualitative comparisons between our method and the baselines. GaussianHair~\cite{luo2024gaussianhair} is overfitted to training lighting conditions, where the model learns baked-in highlights and thus generalizes poorly to novel test lighting. HairInverse~\cite{sun2021human} regresses to an average color of the hair region, lacking the capacity to reconstruct diverse hair colors and highlight patterns. Our method estimates the base appearance in the first stage (\textit{Ours Stage1}) and captures more details and highlights by leveraging the residual module (\textit{Ours Full}).

\input{figs/fig_supp_results_1}
\input{figs/fig_supp_error_ablation}
\input{figs/fig_supp_error_baseline}
\input{figs/fig_supp_relighting}

%% file: figs/fig_supp_results_1.tex
\begin{figure*}[h!]
    \centering
    \includegraphics[width=0.8\linewidth]{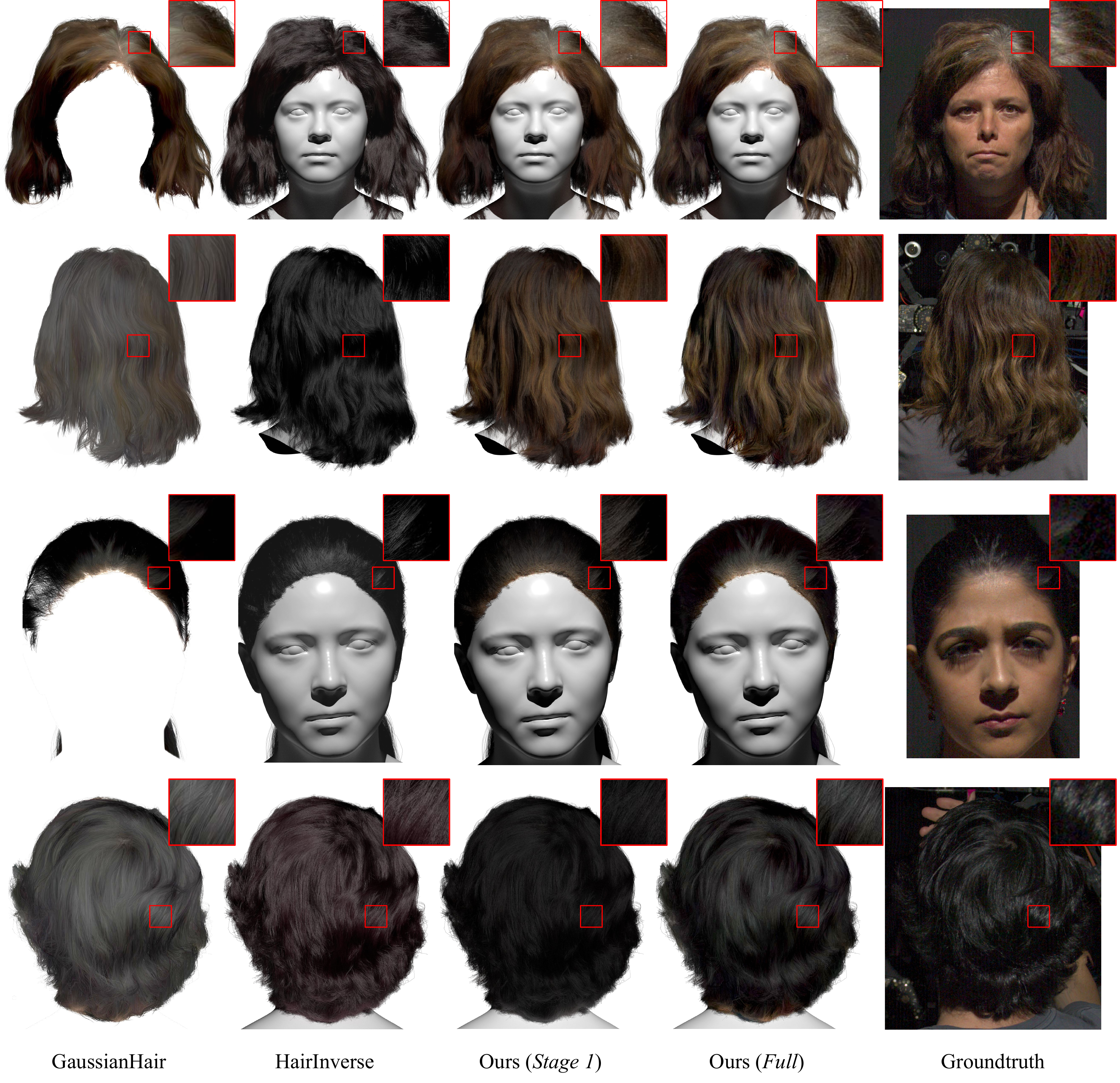}
    \vspace{-5pt}
    \caption{\textbf{More qualitative baseline comparisons} of novel testing views rendering under novel lighting conditions.} 
    \label{fig:results supp 1}
\end{figure*}

%% file: figs/fig_supp_error_ablation.tex
\begin{figure*}[b!]
    \centering
    \includegraphics[width=0.8\linewidth]{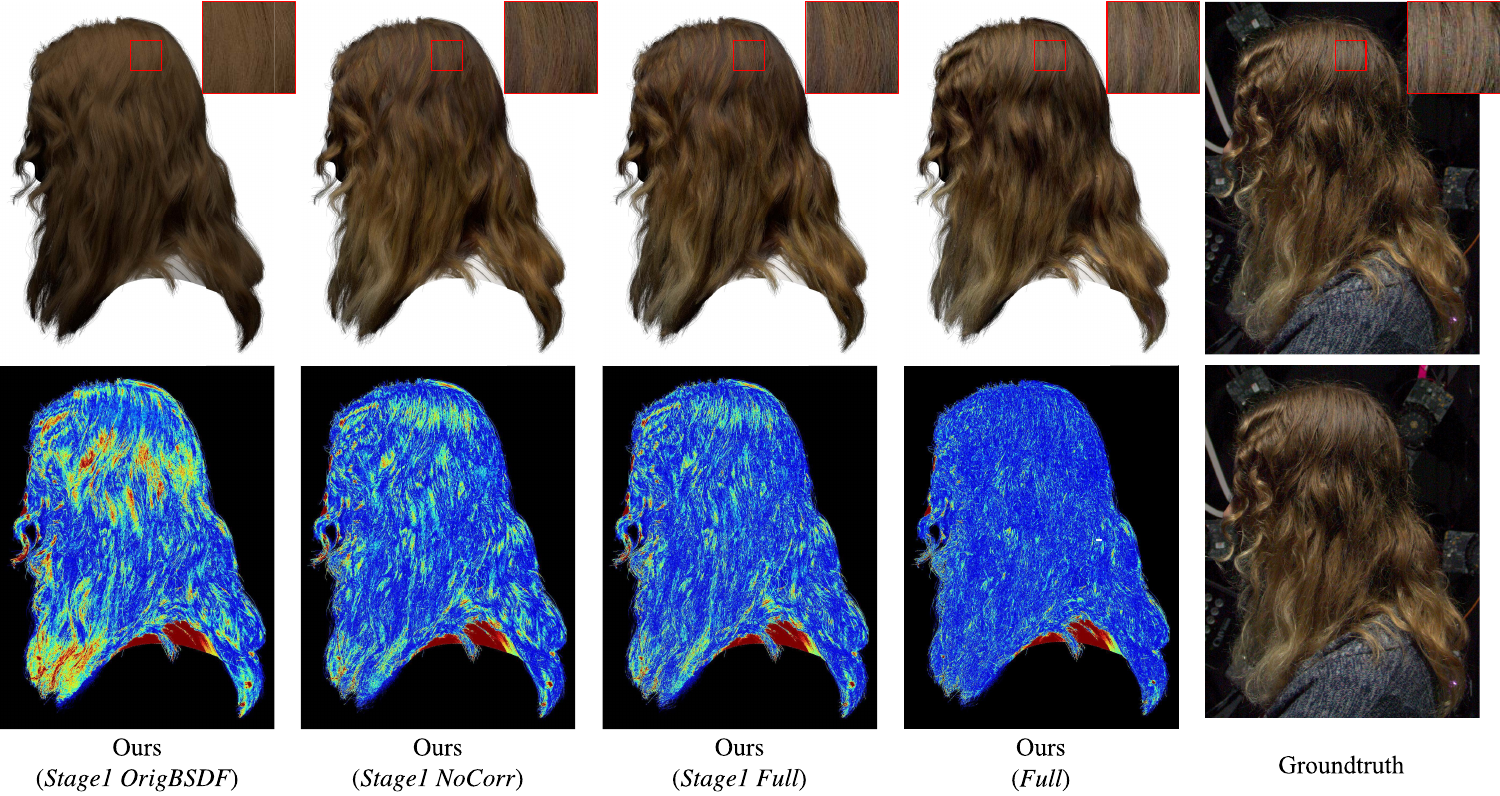}
    \vspace{-5pt}
    \caption{\textbf{Error maps of ablation study} on view synthesis under novel viewpoint and lighting condition.} 
    \label{fig:error_ablation}
\end{figure*}

%% file: figs/fig_supp_error_baseline.tex
\begin{figure*}[h!]
    \centering
    \includegraphics[width=0.8\linewidth]{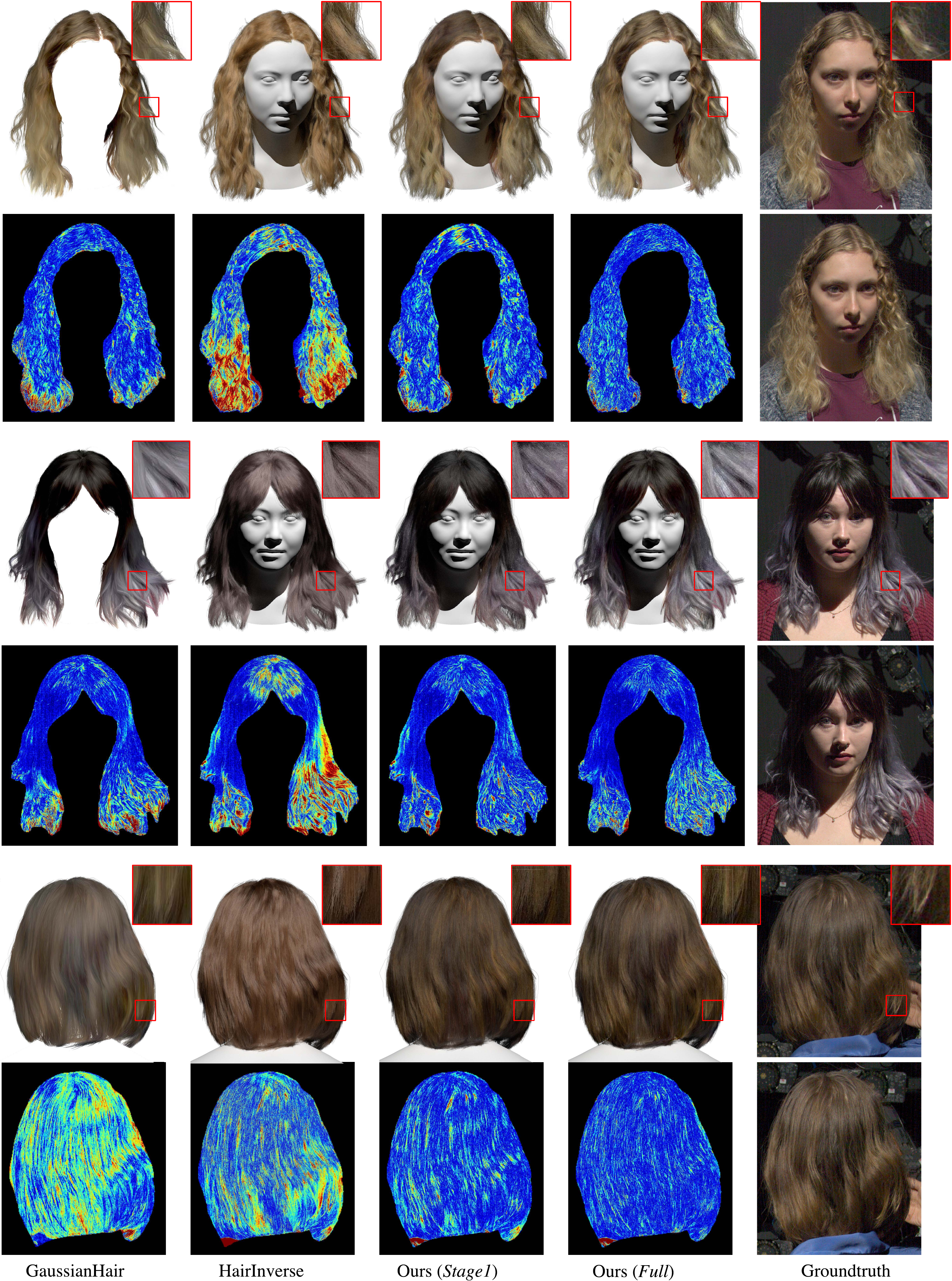}
    \caption{\textbf{Error maps of baseline comparisons} on view synthesis under novel viewpoint and lighting condition.} 
    \label{fig:error_baseline}
\end{figure*}

%% file: figs/fig_supp_relighting.tex
\begin{figure*}[h!]
    \centering
    \includegraphics[width=0.9\linewidth]{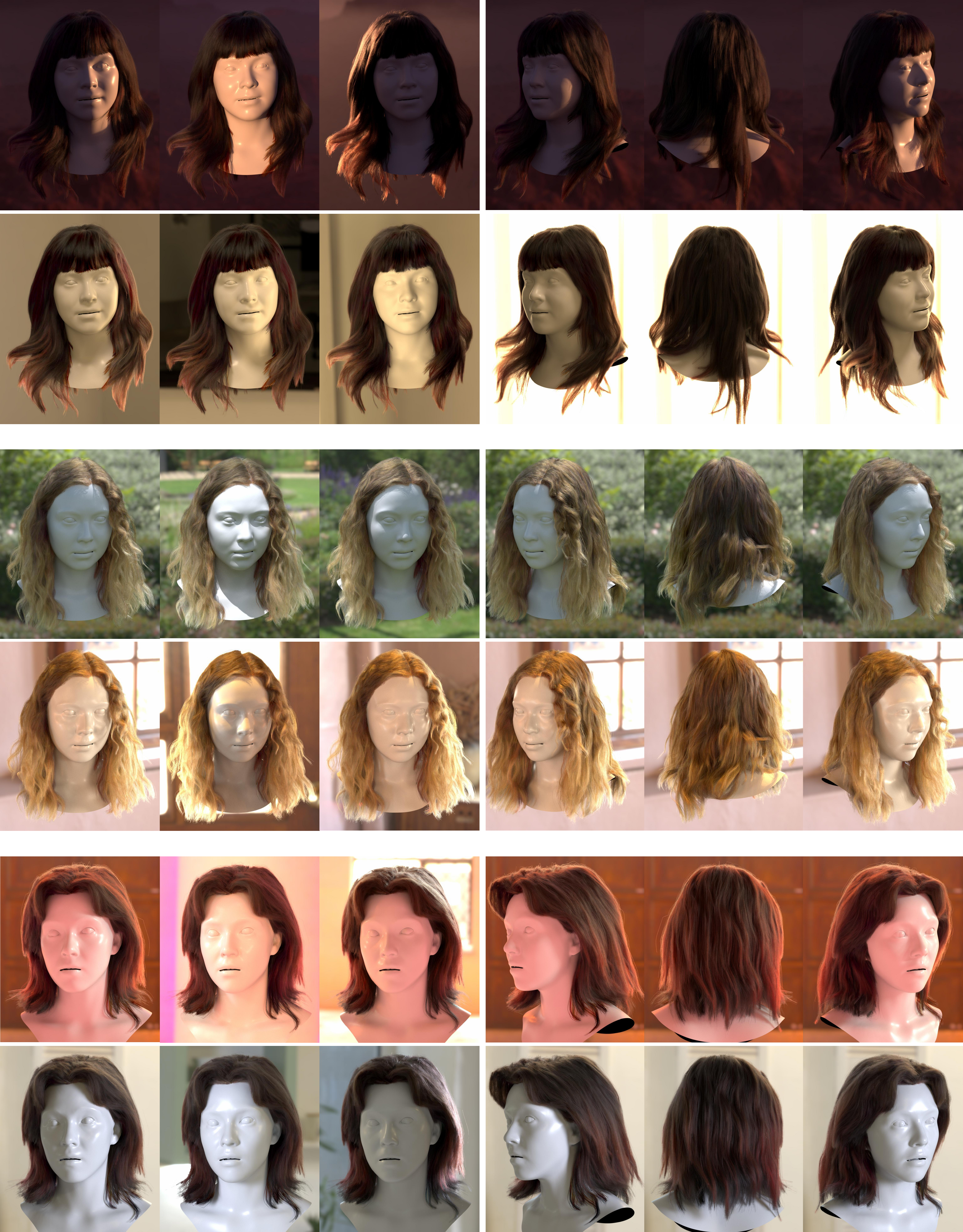}
    \caption{\textbf{More relighting results}. Each row shows relighting results with rotated environmental light (left) and rotated geometry (right).} 
    \label{fig:results_relighting}
\end{figure*}